
\magnification=\magstep1
\parskip 0pt
\def\({[}
\def\){]}
\def\sno{\smallskip\noindent}

\def\lbrk{\break}

\def\endlist{\par}

\hsize 6.00in

\overfullrule 0pt
\centerline{\bf
Bose-Einstein Correlations for Mixed Neutral Mesons
}
\centerline{\bf Harry J. Lipkin}
\smallskip
\centerline{Department of Nuclear Physics}
\centerline{\it Weizmann Institute of Science}
\centerline{Rehovot 76100, Israel}
\centerline{and}
\centerline{School of Physics and Astronomy}
\centerline{Raymond and Beverly Sackler Faculty of Exact Sciences
}
\centerline{\it Tel Aviv University}
\centerline{Tel Aviv, Israel}
\sno
\centerline{To be Published in Physical Review Letters}
\lbrk
\centerline{September 29, 1992}
\vskip 0.2in
\baselineskip 18pt

\abstract
\medskip

Correlations are shown to arise in nonidentical mixed-particle pairs
like $K^o \bar K^o$ when observed in identical decay modes like $K_S
K_S$ in multiparticle final states containing many partial waves.
No enhancement is found in any single partial wave and all partial wave
analyses of the s-wave threshold resonance $a_o$ and $f_o$ should give
the same results for all decay modes. In CP violation experiments where
$B^o - \bar B^o$ pairs are inclusively produced and correlated decays
into $\psi K_S$ and leptonic modes are observed, the CP-violating lepton
asymmetry is enhanced by a factor of two in the kinematic region where
Bose enhancement occurs.
\medskip

The Bose-Einstein correlations observed between identical charged pions
\REF{\Gerson}{G. Goldhaber et al, Phys. Rev. {\bf 120}, (1960) 300
}\refend
or charged kaons
\REF{\Odette}{T. Akesson et al, Phys. Lett. {\bf B155}, (1985) 128
and references cited there
}\refend
in experiments leading to multiparticle final states
are a consequence of quantum mechanics and well understood.
The $K^o$ and  $\bar K^o$ are two different particles and
one might expect no correlations. But detecting neutral kaon pairs
by $2\pi$ decays projects out a final $K_S K_S$ state which does have two
identical bosons (we neglect small $CP$-violation effects).
A straightforward
calculation with all proper interference terms and phases shows an enhancement
in the momentum distribution of the $K_S K_S$ system.
Experiments\REF{\Gideon}{OPAL Collaboration, CERN Preprint
CERN-PPE/92-192  submitted to Phys. Lett.
1992}\refend have now seen a $K_S K_S$ enhancement identical to that found
for the $\pi^+ \pi^+$ system in the region of phase space where the two kaons
have nearly the same momentum.
Whether one calls this Bose enhancement or something else is a matter of
semantics and taste.

In this paper we first show how Bose enhancement arises in production of flavor
neutral meson pairs which are not identical bosons. We then
investigate effects of these Bose correlations on experiments in
hadron spectroscopy and searches for CP violation.

Consider a reaction in which
a final state denoted by $\ket{f(\vec p_\alpha , \vec p_\beta ,\xi)}$
containing a $K^o \bar K^o$ pair with momenta respectively
$\vec p_{\alpha}$ and $\vec p_{\beta}$ is produced inclusively with an
arbitrary
number of assorted other particles,
  $$ \ket{f(\vec p_\alpha , \vec p_\beta ,\xi)} \equiv \ket{K^o(\vec
p_{\alpha}); \bar K^o(\vec p_{\beta}); \xi)} \eqno  (YY1)       $$
where $\xi$ denotes all the other degrees of freedom of the final state.
This final state can be expanded in the basis of the weak interaction
eigenstates $K_L$ and $K_S$,
$$ \ket{f(\vec p_\alpha , \vec p_\beta ,\xi)} =
(1/2)\ket{K_L(\vec p_{\alpha}); K_L(\vec p_{\beta}); \xi}+
(1/2)\ket{K_S(\vec p_{\alpha}); K_S(\vec p_{\beta}); \xi}+ $$
$$ (1/2)\ket{K_S(\vec p_{\alpha}); K_L(\vec p_{\beta}); \xi}-
(1/2)\ket{K_L(\vec p_{\alpha}); K_S(\vec p_{\beta}); \xi}
 \eqno  (YY2)       $$
where we neglect CP-violation.
We now express this state in terms of states with a definite permutation
symmetry  in the two momenta,
  $$ \ket{f(\vec p_\alpha , \vec p_\beta ,\xi)} =
\ket{f_s(\vec p_\alpha , \vec p_\beta ,\xi)} +
\ket{f_a(\vec p_\alpha , \vec p_\beta ,\xi)}  \eqno  (YY3)       $$
where
 $$ \ket{f_s(\vec p_\alpha , \vec p_\beta ,\xi)} \equiv
(1/2) \ket{f(\vec p_\alpha , \vec p_\beta ,\xi)} +
(1/2) \ket{f( \vec p_\beta , \vec p_\alpha ,\xi)}  =
$$
$$ =
(1/2)\ket{K_L(\vec p_{\alpha}); K_L(\vec p_{\beta}); \xi}+
(1/2)\ket{K_S(\vec p_{\alpha}); K_S(\vec p_{\beta}); \xi}
\eqno  (YY4a)       $$
 $$ \ket{f_a(\vec p_\alpha , \vec p_\beta ,\xi)} \equiv
(1/2) \ket{f(\vec p_\alpha , \vec p_\beta ,\xi)} -
(1/2) \ket{f( \vec p_\beta , \vec p_\alpha ,\xi)}  =
$$
$$ =
(1/2)\ket{K_S(\vec p_{\alpha}); K_L(\vec p_{\beta}); \xi}-
(1/2)\ket{K_L(\vec p_{\alpha}); K_S(\vec p_{\beta}); \xi}
\eqno  (YY4b)       $$
We now consider the total probability of observing a kaon pair with momenta
$\vec p_{\alpha}$ and $\vec p_{\beta}$, where we detect the kaons as $K^o$ or
$\bar K^o$ by a strong charge exchange reaction and add the two possible
strangeness states for each momentum. This is proportional to the symmetrized
norm of the state $\ket{f(\vec p_\alpha , \vec p_\beta ,\xi)}$ integrated over
all the other variables $\xi$,
$$ P_{K \bar K}(\vec p_\alpha , \vec p_\beta) = \int d\xi
\langle{f(\vec p_\alpha , \vec p_\beta ,\xi)}\ket{f(\vec p_\alpha ,
\vec p_\beta ,\xi)} +
\langle{f(\vec p_\beta , \vec p_\alpha ,\xi)}
\ket{f(\vec p_\beta , \vec p_\alpha ,\xi)} = $$
$$ \int d\xi
\langle{f_s(\vec p_\alpha , \vec p_\beta ,\xi)}\ket{f_s(\vec p_\alpha ,
\vec p_\beta ,\xi)} +
\langle{f_a(\vec p_\alpha , \vec p_\beta ,\xi)}\ket{f_a(\vec p_\alpha ,
\vec p_\beta ,\xi)}
\eqno  (YY5)       $$

For experiments detecting the kaons as $K_L$ and $K_S$ we obtain
$$ P_{K_L K_L}(\vec p_\alpha , \vec p_\beta) =
P_{K_S K_S}(\vec p_\alpha , \vec p_\beta) = {1\over 2}\cdot
\int d\xi
\langle{f_s(\vec p_\alpha , \vec p_\beta ,\xi)}\ket{f_s(\vec p_\alpha ,
\vec p_\beta ,\xi)}
\eqno  (YY6a)       $$
$$ P_{K_L K_S}(\vec p_\alpha , \vec p_\beta) =
\int d\xi
\langle{f_a(\vec p_\alpha , \vec p_\beta ,\xi)}\ket{f_a(\vec p_\alpha ,
\vec p_\beta ,\xi)}
\eqno  (YY6b)       $$
Thus
$${{ P_{K_L K_L}(\vec p_\alpha , \vec p_\beta)}\over{
P_{K \bar K}(\vec p_\alpha , \vec p_\beta)}} =
{{P_{K_S K_S}(\vec p_\alpha , \vec p_\beta)}\over{
P_{K \bar K}(\vec p_\alpha , \vec p_\beta)}} = $$
$$=
{1\over 2}\cdot{{
\int d\xi
\langle{f_s(\vec p_\alpha , \vec p_\beta ,\xi)}\ket{f_s(\vec p_\alpha ,
\vec p_\beta ,\xi)}}\over{
\int d\xi
\langle{f_s(\vec p_\alpha , \vec p_\beta ,\xi)}\ket{f_s(\vec p_\alpha ,
\vec p_\beta ,\xi)} +
\langle{f_a(\vec p_\alpha , \vec p_\beta ,\xi)}\ket{f_a(\vec p_\alpha ,
\vec p_\beta ,\xi)}}}
\eqno  (YY7a)       $$
$$ {{P_{K_L K_S}(\vec p_\alpha , \vec p_\beta)}\over{
P_{K \bar K}(\vec p_\alpha , \vec p_\beta)}} = $$
$$=
{1\over 2}\cdot{{
\int d\xi
\langle{f_a(\vec p_\alpha , \vec p_\beta ,\xi)}\ket{f_a(\vec p_\alpha ,
\vec p_\beta ,\xi)}}\over{
\int d\xi
\langle{f_s(\vec p_\alpha , \vec p_\beta ,\xi)}\ket{f_s(\vec p_\alpha ,
\vec p_\beta ,\xi)} +
\langle{f_a(\vec p_\alpha , \vec p_\beta ,\xi)}\ket{f_a(\vec p_\alpha ,
\vec p_\beta ,\xi)}}}
\eqno  (YY7b)       $$

When the final state is expressed in terms of a partial wave analysis
the symmetric and antisymmetric states $\ket{f_s}$ and $\ket{f_a}$ project
out respectively the partial waves with even and odd values of the angular
momentum of the kaon pair in their center-of-mass system.
We can therefore draw the following conclusions:
\pointbegin For complicated final states where many partial waves contribute
incoherently, the contributions of even and odd partial waves are roughly equal
outside of the kinematic region of Bose enhancement and the two contributions
to the denominators of eqs. (YY7) are equal. Thus $P_{K_S K_S} =P_{K_L K_L}
= (1/4) P_{K^o \bar K^o}$ and $P_{K_L K_S}= (1/2) P_{K^o \bar K^o}$.
\point In the Bose enhancement region only even partial waves contribute and
the second term in the denominators of eqs. (YY7) vanish. Thus
$P_{K_S K_S} =P_{K_L K_L}= (1/2) P_{K^o \bar K^o}$ and $P_{K_L K_S}= 0$.
This doubling of the $P_{K_S K_S}/P_{K^o \bar K^o}$ ratio is just the Bose
enhancement factor observed in the experiment$^{\Gideon}$.
\point In an experiment in which partial wave analysis separates out the
$s-wave$ using angular distributions, the second term in the denominator
of eq. (YY7) is zero and $P_{K_S K_S} =P_{K_L K_L}= (1/2) P_{K^o \bar K^o}$
over all regions of phase space without
regard to any Bose enhancement.
\endlist
We therefore conclude that in experiments where many partial waves contribute
Bose enhancement is observed experimentally and the
invariant mass spectrum of the kaon pair will show a larger enhancement at
threshold when observed as  $K_S K_S$ or  $K_L K_L$ than when
observed as $K^o \bar K^o$. However, in any experiment which separates
out the s wave the invariant mass spectra of the $K_S K_S$ or $K_L K_L$ pairs
is identical to the $K^o \bar K^o$ spectrum produced in the same experiment.
Thus partial wave analyses using the $K_S K_S$ decay mode give reliable
determinations of the parameters of the $f_o$ and $a_o$ resonances and are not
distorted by Bose enhancement.

We now examine inclusive hadronic production of $B^o-\bar B^o$ pairs
in an incoherent mixture of states which are even and odd under $CP$.
An enhancement by a factor of two of the CP-violating
lepton asymmetry above background can be produced
by Bose-Einstein correlations occur which enhance the even $CP$ states and
suppress odd $CP$. This can be of practical importance in
experiments which search for $CP$ violation by observing lepton
asymmetries in correlated decay modes\REF{\HJL68}{Harry J. Lipkin, Phys.
Rev. {\bf 176}, (1968) 1715}\refend
\REF{\Rosner}{I. Dunietz, J. Hauser and J. L. Rosner, Phys.
Rev. {\bf D35} (1987) 2166}\refend
\REF{\Bernabeu}{J. Bernabeu, F. J. Botella and J. Roldan, Phys.
Lett. {\bf B211}, (1988) 226}\refend, where one $B$ decays into a CP
eigenstate and the other into a leptonic mode. In an experiment with no
time measurement the result gives only a time integral of the
lepton asymmetry. Only the asymmetry from the even $CP$ state
survives the time integration; the asymmetry from the odd $CP$ state
averages to zero and gives a symmetric background
\REF{\Bigi}{I. I. Bigi and A. I Sanda, Nuc. Phys.
{\bf B281}, (1987) 41.
}\refend.
Thus an enhancement
by a factor of two of the lepton asymmetry above background
should be observed in the kinematic region where the Bose-Einstein
correlations occur.

The effect is most simply seen in the quasispin formulation
\REF{\Bspin}{Harry J. Lipkin,
Argonne report no
ANL-HEP-PR-88-66 (to be published)
}\refend
where the $B^o$ and $\bar B^o$ are classified in a doublet of an SU(2)
quasispin algebra
\REF{\TDLee}{T. D. Lee and C. S. Wu, Ann. Rev. Nuc. Sci. 16, (1966) 511
}\refend
and a generalized Pauli principle can be defined which requires
the wave function for a two-meson system to be even under permutations
in both space and quasispin
\REF{\PEPRspin}{Harry J. Lipkin,
Physics Letters B219, (1989) 474}\refend.
The odd-C state of the two-meson system is
a quasispin singlet, odd under quasispin permutations, while
the three even-C states constitute a quasispin triplet even under
quasispin permutations. Thus only the even-C quasispin-triplet states
are allowed to be in the same spatial quantum state and to contribute to
the total wave function in the kinematic region of a multiparticle
process where Bose-Einstein correlations occur. The odd-C quasispin
singlet state is forbidden for two mesons in the same spatial quantum
state and should be suppressed rather than enhanced in this
kinematic region.

This point can also be seen by noting the form of a $B^o \bar B^o$
pair wave function with even and odd values of $CP$
$$ \ket{(B \bar B)_{ab}}^{\pm} = (1/\sqrt 2)(
\ket {B^o_a \bar B^o_b} \pm \ket {B^o_b \bar B^o_a})
\eqno (YY8)  $$
where $a$ and $b$ are the quantum numbers; e.g. momentum, which label
the spatial state of the meson. When a=b; i.e.
the two mesons are in the same quantum state or for example have the same
momentum, the odd $CP$ wave function vanishes and the norm of the even
$CP$ wave function increases.

We now demonstrate the effect explicitly using detailed wave functions
for the case of a $CP$ violation experiment measuring a lepton asymmetry
in coincidence with a $\psi-K_S$ decay of the other meson.
To show explicitly how factors of two and phases arise in a coherent
calculation without skipping steps we reproduce some well-known equations
without claiming originality.
Consider the case of a $B^o - \bar B^o$ pair created in a hadronic
interaction together with other hadrons. The Bose enhancement will be
observed if the two mesons are ``identical" and not observed if they are
not identical. At first sight the $B^o$ and $\bar B^o$ are not identical.
But the correlated wave function can be expressed in a basis where there
are terms in which the two particles are identical.

Let us choose the basis of CP eigenstates denoted by $B_1$ and $B_2$,
where $B_2$ is the odd-CP state allowed to decay into the odd-CP eigenstate
$K_S \psi$.
$$
\vbox{\eqalignno{
\ket{B_1}    &= (1/\sqrt 2)\cdot (\ket{B^o} - \ket{\bar B^o})
&(YY9a)\cr
\ket{B_2}    &= (1/\sqrt 2)\cdot (\ket{B^o} + \ket{\bar B^o}).
&(YY9b)\cr}} $$
A $B^o - \bar B^o$ state with odd angular momentum in its center of mass
is a $B_1 B_2$ state of two different particles which shows no Bose
enhancement. A $B^o - \bar B^o$ state with even angular momentum in its center
of mass is a linear combination of $B_1 B_1$ and $B_2 B_2$. These states of two
identical particles can show the Bose enhancement.

We now construct the transition matrices explicitly.
We first consider the matrix element for the transition
$$ \bra {L_a^{\pm} (K_S \psi)_b} T \ket {B^a_2 B^b_2}=
\bra {L_b^{\pm} (K_S \psi)_a} T \ket {B^a_2 B^b_2}=
\bra {K_S \psi} T \ket {B_2}\cdot  \bra {L_{\pm}} T \ket {B_2}
\eqno (YY10a)  $$
$$ \bra {L_a^{\pm} (K_S \psi)_b} T \ket {B^a_1 B^b_1}=
\bra {L_b^{\pm} (K_S \psi)_a} T \ket {B^a_1 B^b_1}=0
\eqno (YY10b)  $$
where $a$ and $b$ denote the spatial quantum numbers of the particular
state,
$T$ denotes the transition operator and $L_{\pm}$ denotes any given
leptonic decay mode. The transition from the initial
state $     {B^a_1 B^b_1}$ into the $      {L^{\pm} K_S \psi} $ decay
mode vanishes because $ B_1 \rightarrow K_S \psi $ is forbidden.
The total transition probability from the initial
state $\ket {B^a_2 B^b_2}$ into the $ \ket {L^{\pm} K_S \psi} $ decay
mode is proportional to the square of the transition matrix element
$$ W(B^a_2 B^b_2 \rightarrow L^{\pm} K_S \psi) =
|\bra {L_a^{\pm} (K_S \psi)_b} T \ket {B^a_2 B^b_2} +
\bra {L_b^{\pm} (K_S \psi)_a} T \ket {B^a_2 B^b_2} |^2=$$
$$ = 2(1+\delta_{ab})
|\bra {K_S \psi} T \ket {B_2}\cdot  \bra {L_{\pm}} T \ket {B_2}|^2
\eqno (YY10c)  $$
where the factor 2 is seen to arise from the fact that there are two
possible transitions which are incoherent.
Either of the two mesons can decay into $K_S \psi$ with the other
decaying into leptons, but the two final states have different spatial
quantum numbers when $a \not= b$. The states are therefore not coherent
and the transition matrix elements are squared before adding to give the
total transition probability. However when $a=b$ the two final states are
identical, the two transition matrix elements are coherent, and the two
terms are added before squaring to give the enhancement.
This additional factor 2 is the standard factor familiar in coherent
``stimulated emission" arising from the fact that the two identical
bosons are in the same quantum state.

We now examine the effect of these ``Bose"-type correlations on the
explicit time dependence of the $CP$ asymmetry observed in experiments.
Let $\ket {B^o(t)}$ and $\ket {\bar B^o(t)}$ denote the quantum state at
time $t$ in the
$B^o$ - $\bar B^o$
degree of freedom of two states which are respectively pure
$B^o$ and pure $\bar B^o$
at time $t=0$.
These two states will oscillate in time between the states
$B^o$ and $\bar B^o$. The explicit form of these oscillations is
$^{\Bigi,\PEPRspin}$
$$
\vbox{\eqalignno{
\ket{B^o(t)} &=
e^{-{\Gamma\over 2} t}
 \cdot\{\cos
({{\omega t}\over 2})
\ket{B^o}
- ie^{i\theta}
\sin ({{\omega t}\over 2})
\ket{\bar B^o}
\}
&(YY11a)\cr
 \ket{\bar B^o(t)} &=
e^{-{\Gamma\over 2} t}
 \cdot\{\cos
({{\omega t}\over 2})
\ket{\bar B^o}
- ie^{-i\theta}
\sin ({{\omega t}\over 2})
\ket{B^o}
\}
&(YY11b)\cr}} $$
where $\Gamma$ denotes the decay width, $\omega$ is the mass
difference between the two eigenstates and $\theta$ is a parameter
expressing the CP violation$^{\PEPRspin}$.

We now consider the matrix element for the transitions
$$ \bra {L^+_a (K_S \psi)_b} T \ket {B^o_a(t_1) \bar B^o_b(t_2)}=
$$
$$ = \bra {K_S \psi} T \ket {B_2}\cdot  \bra {L^+} T \ket {B^o}
\langle B_2 \ket{\bar B^o(t_2)}\cdot \langle B^o \ket{B^o(t_1)}
\eqno (YY12a)  $$
$$ \bra {L^+_b (K_S \psi)_a} T \ket {B^o_a(t_2) \bar B^o_b(t_1)}=
$$
$$ = \bra {K_S \psi} T \ket {B_2}\cdot  \bra {L^+} T \ket {B^o}
\langle B^o \ket{\bar B^o(t_1)}\cdot
\langle B_2 \ket{B^o(t_2)}
\eqno (YY12b)  $$
$$ \bra {L^-_a (K_S \psi)_b} T \ket {B^o_a(t_1) \bar B^o_b(t_2)}=
$$
$$ =\bra {K_S \psi} T \ket {B_2}\cdot  \bra {L^-} T \ket {\bar B^o}
\langle B_2 \ket{\bar B^o(t_2)}\cdot \langle \bar B^o \ket{B^o(t_1)}
\eqno (YY13a)  $$
$$ \bra {L^-_b (K_S \psi)_a} T \ket {B^o_a(t_2) \bar B^o_b(t_1)}=
$$
$$ = \bra {K_S \psi} T \ket {B_2}\cdot  \bra {L^-} T \ket {\bar B^o}
\langle \bar B^o \ket{\bar B^o(t_1)}\cdot
\langle B_2 \ket{B^o(t_2)}
\eqno (YY13b)  $$
where $a$ and $b$ denote the spatial quantum numbers of the particular
state, $T$ denotes the transition operator,
$L^{+}$ denotes any given
leptonic decay mode with a positive lepton and which therefore must come
from a $B^o$ decay and not from a $\bar B^o$ decay, $L^{-}$ denotes any
given
leptonic decay mode with a negative lepton and which therefore must come
from a $\bar B^o$ decay and not from a $B^o $decay,
$t_1$ is the time of the leptonic decay and
$t_2$ is the time of the $ K_S \psi $ decay.
The total transition probability from the initial
state $\ket {B^o_a(t_1) \bar B^o_b(t_2)}$ into the
$L^{\pm}$ and $K_S \psi $ decay modes at times $t_1$ and $t_2$
respectively is proportional to the incoherent sum of the
squares of the transition matrix elements (YY12) and (YY13) respectively
if $a \not= b$.
For the case where $a=b$
the two transitions (YY12a) and (YY12b) are coherent and also
(YY13a) and (YY13b) are coherent and there are interference terms.
In an experiment with no time measurement these transition probabilities
must be integrated over $t_1$ and $t_2$ to obtain the experimental result.
Substituting eqs. (YY11) into the expressions (YY12) and (YY13)
for the transition matrix elements, squaring and integrating
gives the lepton asymmetry,
$$
{{\int dt_1 \int dt_2 e^{-2\Gamma \bar t} \{
|\bra {L^+_a (K_S \psi)_b} T \ket {B^o_a(t_1) \bar B^o_b(t_2)}|^2
 -|\bra {L^-_a (K_S \psi)_b} T \ket {B^o_a(t_1) \bar B^o_b(t_2)}|^2
\}}
\over {\int dt_1 \int dt_2 e^{-2\Gamma \bar t} \{
|\bra {L^+_a (K_S \psi)_b} T \ket {B^o_a(t_1) \bar B^o_b(t_2)}|^2
 +|\bra {L^-_a (K_S \psi)_b} T \ket {B^o_a(t_1) \bar B^o_b(t_2)}|^2
\}}} = $$
$$ = -
(1 + \delta_{ab})I_cI_s \sin\theta \eqno(YY14) $$

where
$$ I_c =  \Gamma \int dt e^{-\Gamma t} \cos ({{\omega t}}) \cdot
\eqno(YY15a) $$
$$ I_s = \Gamma \int dt e^{-\Gamma t} \sin ({{\omega t}})
\eqno(YY15b) $$

The lepton asymmetry is seen to be
enhanced by a factor of 2 when the $B^o$ and the $\bar B^o$ are
in the same quantum state.

In conclusion we note that the following example of identical particle effects
in inclusive two-pion decays of any $\Upsilon$ charmonium state may clarify
the basic physics of these Bose enhancements. From isospin invariance we obtain
the relations

$$ {{BR[\Upsilon \rightarrow \pi^+(\vec p_{\alpha})\pi^+(\vec
p_{\alpha})X^{--}]
}\over
{BR[\Upsilon \rightarrow \pi^+(\vec p_{\alpha})\pi^o(\vec p_{\alpha})X^{-}]
}} = 1
\eqno(YY16a) $$
$$ {{BR[\Upsilon \rightarrow (\pi^+ \pi^+)_{s ~ wave} X^{--}]
}\over {BR[\Upsilon \rightarrow (\pi^+ \pi^o)_{s ~ wave} X^{-}]
}} = 1
\eqno(YY16b) $$
$$
{{BR[\Upsilon \rightarrow \pi^+(\vec p_{\alpha})\pi^+(\vec p_{\beta})X^{--}]
}\over
{BR[\Upsilon \rightarrow \pi^+(\vec p_{\alpha})\pi^o(\vec p_{\beta})X^{-}]
+ BR[\Upsilon \rightarrow \pi^o(\vec p_{\alpha})\pi^+(\vec p_{\beta})X^{-}]
}} \leq 1
\eqno(YY16c) $$
The equalities (YY16a) and (YY16b) follow from the observation that the
$\pi^+ \pi^+$ and $\pi^+ \pi^o$ states are two members of the same I=2 isospin
multiplet when the two pions either have the same momentum or are in a relative
s wave. The inclusive production of two members of the same isospin multiplet
from an initial I=0 state must be equal from isospin invariance. However, in
the general case of arbitrary different uncorrelated momenta (YY16c)
an additional I=1 $\pi^+ \pi^o$ contribution absent in $\pi^+ \pi^+$
arises from the states where the two pions are in the odd partial waves
forbidden for identical particles.

That the inequality (YY16c) becomes an equality when the two pion momenta are
equal can be viewed as Bose enhancement for the two identical particles if one
wishes and may be a large effect since an appreciable I=1 contribution is
expected in the general case. It can also be viewed as the
suppression of the contribution from odd partial waves in the kinematic region
of Bose enhancement. However, there is no enhancement of the identical
$\pi^+ \pi^+$ pairs over the nonidentical $\pi^+ \pi^o$ after the s-wave is
projected out (YY16b).

Discussions with Gideon Alexander are gratefully acknowledged, and in
particular his suggestion that Bose correlations in mixed neutral meson
states might be of interest.
\refout
\end